\documentclass[useAMS,usenatbib]{mnras}

\usepackage{graphicx}	

\title[Formation of dSph including their SFH]{A possible formation
  scenario for dwarf spheroidal galaxies - III. Adding star formation
  histories to the fiducial model}  

\author[Alarc\'on et al.]{A.G. Alarc\'on Jara$^{1}$\thanks{E-mail:
    alexralarconj@udec.cl}, 
  M. Fellhauer$^{1}$,
  D.R. Matus Carrillo$^{1}$,
  P. Assmann$^{1}$,
  \newauthor F. Urrutia Zapata$^{1}$,
  J. Hazeldine$^{1}$ 
  and C.A. Aravena$^{1}$\\
  $^{1}$Departamento de Astronom\'ia, Universidad de Concepcion, Casilla
  160-C, 3349001 Concepci\'on, Chile}

\begin{document}

\label{firstpage}

\pagerange{\pageref{firstpage}--\pageref{lastpage}}\pubyear{2017}

\maketitle

\begin{abstract}
  Dwarf spheroidal galaxies are regarded as the basic building blocks 
  in the formation of larger galaxies and are believed to be the most
  dark matter dominated systems known in the Universe.  There are
  several models that attempt to explain their formation and
  evolution, but they have problems to model the formation of isolated
  dwarf spheroidal galaxies.  Here, we will explain a possible
  formation scenario in which star clusters form inside the dark
  matter halo of a dwarf spheroidal galaxy.  Those star clusters
  suffer from low star formation efficiency and dissolve while
  orbiting inside the dark matter halo.  Thereby, they build the faint
  luminous components that we observe in dwarf spheroidal galaxies.
  In this paper we study this model by adding different star formation
  histories to the simulations to compare the results with our previous
  work and observational data to show that we can explain the
  formation of dwarf spheroidal galaxies.  
\end{abstract}

\begin{keywords}
  methods: numerical --- galaxies: dwarfs --- galaxies: star clusters:
  general --- galaxies: star formation --- galaxies: structures ---
  cosmology: dark matter  
\end{keywords}

\section{Introduction}
\label{sec:intro}

The Local Group consists of about 80 galaxies, discovered so far
\citep[e.g.][]{McConnachie2012}.  Most of them are dwarf galaxies,
orbiting the two large galaxies (Milky Way (MW) and Andromeda (M31)).
Their classifications range from dwarf disc galaxies via dwarf
irregulars, dwarf ellipticals and compact ellipticals to dwarf
spheroidal (dSph) galaxies, which are the majority of the known
dwarfs. 

Dwarf spheroidal galaxies exhibit high mass-to-light (M/L) ratios and
are believed
to be the most dark matter (DM) dominated stellar systems known
\citep[e.g.][]{Mateo1998,Belokurov2007,Walker2009}.  They are among
the oldest structures and are by far the most numerous galaxies in the
Universe, however, due to their intrinsic faintness the study of dSph
galaxies has been very difficult.  In the standard hierarchical galaxy
formation models, dwarf galaxies are the elemental systems in the
Universe; larger galaxies are formed from smaller objects like dwarf
galaxies through major and/or minor mergers
\citep{Kauffmann1993,Cole1994}.  Thus, it is important to study these 
galaxies to understand the formation and evolution of normal sized
galaxies. 

DSph have a low stellar content and are poor in, or entirely devoid
of gas.  They are widely thought as the smallest cosmological
structures containing DM in the Universe
\citep{Mateo1998,Lokas2009,Walker2009}.  The classical dSph galaxies
of the MW are characterized by absolute magnitudes in the range $-13
\leq M_{\rm V} \leq -9$ \citep{Mateo1998}.  With the discovery of the faint
and ultra-faint dSph galaxies the range of luminosities is extended to
objects with only a few hundred solar luminosities
\citep[e.g. Segue~1,][]{Belokurov2007}.  We expect them to form in a similar
way than the classical dSph, but we exclude them in this study.
The total estimated masses of the classical dSphs, considering the stars and
the DM
halo is of the order of $10^7-10^8$~M$_{\odot}$.  They exhibit high velocity
dispersions \citep[e.g.][]{Simon2007,Koch2009}; in the range of $5$-$12$
km\,s$^{-1}$ and the velocity dispersion remains approximately
constant with distance from the center of the galaxy
\citep{Kleyna2002,Kleyna2003,Kleyna2004,Munoz2005,Munoz2006,Simon2007,Walker2007}.  

There are several models that attempt to explain the origin of dSph
galaxies by considering different mechanisms.  Some of them are based
on tidal and ram-pressure stripping \citep{Gnedin1999,Mayer2007}.  In
these models, the dSph galaxies are formed due to the interaction
between a rotationally supported dwarf irregular galaxy and a MW-sized
host galaxy.  These models show that dSph galaxies tend to appear
near massive galaxies, but they do not explain the presence of distant
isolated dSph galaxies, such as Tucana and Cetus. 

The model proposed by \citet{DOnghia2009} considers a mechanism known
as resonant stripping and can be used to explain the formation of
isolated dSph galaxies.  Basically, these objects are formed after
encounters between dwarf disc galaxies, in a process driven by
gravitational resonances. 

Here, we study the dissolving star cluster scenario proposed by
\citet{Assmann2013a,Assmann2013b}.  According to this model a dSph
galaxy is formed by the fusion and dissolution of several star
clusters (SCs) with low star formation efficiency (SFE), formed within
a small DM halo, which later hosts the dSph galaxy.  This model does
not need the gravitational interaction with other galaxies to explain
the formation of dSphs, therefore, we can use it to explain the
formation of isolated dSph galaxies.  

This model is based on the assumption that stars never form in
isolation but in hierarchical structures \citep[i.e.\ SCs,
e.g.][]{Lada2003,Lada2010}.  Star formation events range from
slowly forming stars in small clusters and associations to intense
starbursts, in gas-rich environments, typically producing a few to few
hundred young SCs, within a region of just a few hundred pc
\citep[e.g.][]{Withmore1999}. 

The SCs form embedded inside a molecular gas cloud and eventually
expel their remaining gas via various feedback processes such as
stellar winds, ultraviolet (UV) radiation and finally the onset of
supernovae 
\citep[e.g.][]{Goodwin1997a,Goodwin1997b,Boily2003a,Boily2003b,Parmentier2008,Bonnell2011,Smith2011a,Smith2011b}.
If the star formation efficiency (SFE) is low then the SCs are not
able to form bound entities and instead will disperse their stars and
dissolve. 

The dynamical evolution of these SCs, i.e.\ their dissolution due to
gas expulsion, thereby spreading their stars in the central area of
the host DM halo, explains the formation of classical dSph galaxies.
As an extension of the work of \citet{Assmann2013a,Assmann2013b}, in
which all SCs were formed at the beginning of the simulations, in
our scenario we add star formation histories (SFH) to the models
placing SCs into the simulation at different moments in time.  The SCs
form with low SFE and thus, are designed to dissolve inside the DM
halo to form the luminous component of the dSph galaxy.  We follow the
evolution of the SCs within the DM halo for $10$~Gyr, and then we
measure the properties of the final object. 

To mimic the SFHs of dSph galaxies we follow the observational data of
\citet[][see their figure 12]{Weisz2014}.  They show that the SFH of
classical dSphs range from intense star bursts, where almost all stars
of the galaxies were formed at the same time, as Sculptor for example,
until galaxies with a constant star formation rate like Fornax or Leo
I. 

In our previous studies all the SCs were formed at beginning of the
simulation, this mimics a dSph with a SFH with a strong starburst.  In
this project we will show that we can explain the properties of
classical dSph galaxies even using different SFHs. 

In the next Section we describe the setup of our simulations.  We
present the results of our models in Section~\ref{sec:results}, we 
discuss them in Section~\ref{sec:disc} and in the final Section we 
give our conclusions. 

\section{Setup}
\label{sec:setup}

\begin{figure}
  \includegraphics[width=8.5cm]{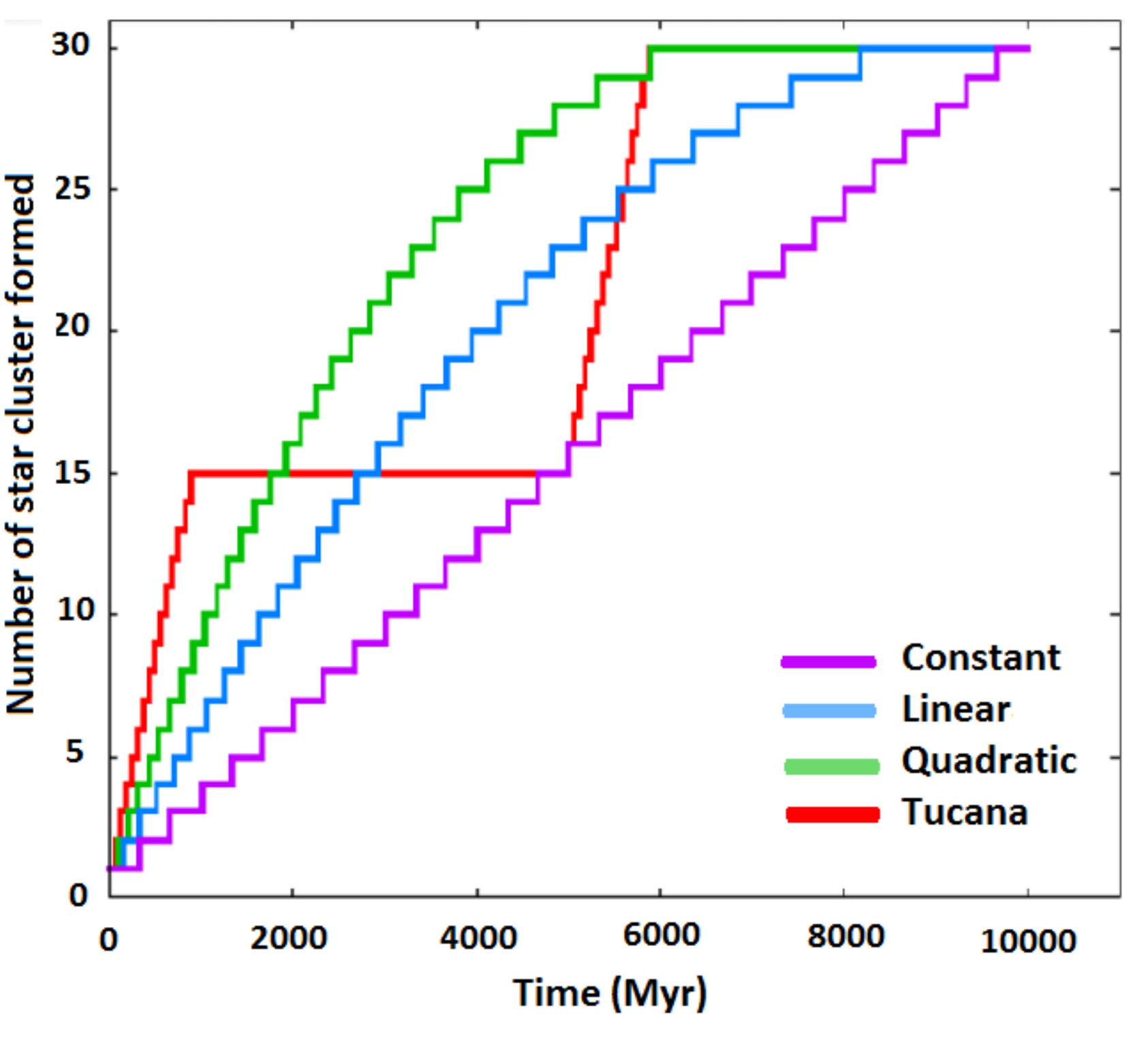}
  \caption{Star formation histories used in this project, based on
    the star formation histories of classical dSph galaxies, taken from the
    analysis of the stars in the main sequence described in
    \citet{Weisz2014}}.  
  \label{fig:sfh}
\end{figure}

\begin{table*}
  \caption{Initial conditions and results of our simulations with a
    cusped DM profile: The first column is the number assigned to the
    simulation, second column is the dark matter profile, third column
    is the SFH,
    fourth column is the star formation efficiency, fifth, sixth and
    seventh column are the fitted parameters of the Sersic profile,
    eighth column is the clumpiness value and the last column is the
    number of surviving star clusters.}
  \label{tab:iniresn}
  \begin{tabular}{|l|l|l|l|l|l|l|l|l|}
    \hline
    Nr. & N/P & SFH & SFE & $\Sigma_{\rm eff}$ &  $R_{\rm eff}$ & n &  C & Surv. \\         
      & & & & $\left[\frac{{\rm M}_{\odot}}{{\rm pc}^{2}}\right]$ &
    [pc] & & & SC(s) \\ \hline
   
   1 & N  &     cte &   0.2                &   0.808 $\pm$ 0.033        &   234 $\pm$ 7        &   0.696 $\pm$ 0.026    &   0.046       &                 0 \\          
   2& N  &     cte &   0.2                &   0.767 $\pm$ 0.036       &   234 $\pm$ 8         &   0.784 $\pm$ 0.030     &   0.046        &                0 \\ 
   3&  N  &     cte &   0.2                &   0.574 $\pm$ 0.033        &   292 $\pm$ 12       &   0.660 $\pm$ 0.034    &   0.048        &                 0 \\ 
   4&  N  &     cte &   0.2               &   0.699 $\pm$ 0.039        &   249 $\pm$ 11       &   1.020 $\pm$ 0.039     &   0.035        &               0 \\  \hline        
   &  N  &     ALL &   0.2               &   0.712 $\pm$  0.102         &  252 $\pm$ 27.42       &   790 $\pm$ 0.161   &   0.043 $\pm$ 0.005       &                0 \\ \hline
   
   5 & N  &     lin &   0.2                &   0.891 $\pm$ 0.032        &   227 $\pm$ 6       &   0.548 $\pm$ 0.022     &   0.028       &                 0 \\         
   6& N  &     lin &   0.2                &   0.796 $\pm$ 0.034       &   229 $\pm$ 7         &   0.748 $\pm$ 0.028     &   0.019         &               0 \\        
   7&  N  &     lin &   0.2               &   0.615 $\pm$ 0.017        &   288 $\pm$ 6        &   0.554 $\pm$ 0.016     &   0.036             &           0 \\        
   8& N  &     lin &   0.2                &   0.848 $\pm$ 0.042        &   226 $\pm$ 8        &   0.850 $\pm$ 0.033     &   0.020              &          0 \\ \hline           
    & N  &      ALL &  0.2                &   0.787 $\pm$ 0.121         &  242.5 $\pm$ 30.35     &   0.675 $\pm$ 0.149  &   0.025 $\pm$ 0.007       &                0  \\ \hline        

   9 & N  &     quad &   0.2               &   0.765 $\pm$ 0.027        &   246 $\pm$ 6        &   0.645 $\pm$ 0.022     &   0.019       &                 0 \\
   10& N  &     quad &   0.2               &   0.829 $\pm$ 0.032        &   226 $\pm$ 6         &   0.703 $\pm$ 0.025     &   0.014          &              0 \\
   11& N  &     quad &   0.2               &   0.640 $\pm$ 0.022        &   280 $\pm$ 7        &   0.553 $\pm$ 0.020     &   0.025              &          0 \\
   12&  N  &     quad &   0.2              &   0.842 $\pm$ 0.041        &   229 $\pm$ 8        &   0.790 $\pm$ 0.032     &   0.016             &           0 \\ \hline    
     &  N  &     ALL  &   0.2              &   0.769 $\pm$ 0.092        &   245.25 $\pm$ 24.78  &   0.672 $\pm$ 0.099     &   0.018 $\pm$ 0.004       &                0 \\ \hline                      

   13 & N  &     tuc &   0.2                &   0.778 $\pm$ 0.029       &   242 $\pm$ 6        &   0.646 $\pm$ 0.023     &   0.019          &              0 \\       
   14&  N  &     tuc &   0.2               &   0.869 $\pm$ 0.035        &   219 $\pm$ 6         &   0.690 $\pm$ 0.025     &   0.018         &              0 \\       
   15&  N  &     tuc &   0.2               &   0.650 $\pm$ 0.022        &   275 $\pm$ 7        &   0.559 $\pm$ 0.020    &   0.037           &              0 \\          
   16&  N  &     tuc &   0.2               &   0.921 $\pm$ 0.038        &   218 $\pm$ 6        &   0.725 $\pm$ 0.026    &   0.015           &              0 \\ \hline         
     &  N  &     ALL &   0.2               &   0.804 $\pm$ 0.118        &   238.5 $\pm$ 26.73  &   0.655 $\pm$ 0.071     &  0.022 $\pm$ 0.009       &              0 \\ \hline                                 
 
    17& N  &     cte &   0.3                &   0.904 $\pm$ 0.039        &   220 $\pm$ 7        &   0.670 $\pm$ 0.027    &   0.071              &          2 \\       
    18& N  &     cte &   0.3                &   0.723 $\pm$ 0.056        &   238 $\pm$ 14       &   0.93 $\pm$ 0.052     &   0.055              &          1 \\        
    19& N  &     cte &   0.3                &   0.716 $\pm$ 0.070        &   257 $\pm$ 18       &   0.543 $\pm$ 0.058    &   0.061              &          1 \\       
    20 &N  &     cte &  0.3                 &   0.845 $\pm$ 0.053        &   222 $\pm$ 10       &   0.95 $\pm$ 0.043     &   0.053              &          3 \\ \hline          
       &N  &     ALL &  0.3                 &   0.797 $\pm$ 0.092        &   234.25 $\pm$ 17.17 &   0.773 $\pm$ 0.199    &   0.060 $\pm$ 0.008  &          1.75 $\pm$ 0.957 \\ \hline     

    21& N  &     lin &   0.3                &   0.828 $\pm$ 0.034        &   233 $\pm$ 7        &   0.672 $\pm$ 0.026    &   0.038              &          0 \\      
    22& N  &     lin &   0.3                &   0.746 $\pm$ 0.007        &   238 $\pm$ 1        &   0.83 $\pm$ 0.011     &   0.019              &          0 \\       
    23& N  &     lin &   0.3                &   0.678 $\pm$ 0.025        &   269 $\pm$ 7        &   0.569 $\pm$ 0.022    &   0.045              &          0 \\    
    24& N  &     lin &   0.3                &   0.512 $\pm$ 0.130        &   306 $\pm$ 60       &   1.30 $\pm$ 0.183     &   0.021              &          1 \\\hline             
      & N  &     ALL &   0.3                &   0.691 $\pm$ 0.134        &   261.5 $\pm$ 33.66  &   0.842 $\pm$ 0.323    &   0.030 $\pm$ 0.012  &          0.25 $\pm$ 0.5 \\\hline      

    25& N  &     quad &   0.3               &   0.712 $\pm$ 0.012        &   254 $\pm$ 2        &   0.748 $\pm$ 0.017    &   0.039              &          0 \\      
    26& N  &     quad &   0.3               &   0.878 $\pm$ 0.039        &   217 $\pm$ 7        &   0.729 $\pm$ 0.028    &   0.016              &          0 \\        
    27& N  &     quad &   0.3               &   0.630 $\pm$ 0.021        &   279 $\pm$ 7        &   0.623 $\pm$ 0.020    &   0.031              &          0 \\        
    28& N  &     quad &   0.3               &   0.898 $\pm$ 0.043        &   219 $\pm$ 8        &   0.830 $\pm$ 0.032    &   0.018              &          0 \\ \hline         
      & N  &     ALL  &   0.3               &   0.779 $\pm$ 0.129        &   242.25 $\pm$ 29.81 &   0.732 $\pm$ 0.085    &   0.026 $\pm$ 0.010  &          0 \\ \hline       

    29& N  &     tuc &   0.3                &   0.730 $\pm$ 0.030        &   250 $\pm$ 8        &   0.749 $\pm$ 0.026    &   0.029              &          0 \\       
    30& N  &     tuc &   0.3                &   0.773 $\pm$ 0.008        &   235 $\pm$ 1        &   0.786 $\pm$ 0.012    &   0.022              &          0 \\       
    31& N  &     tuc &   0.3                &   0.605 $\pm$ 0.022        &   285 $\pm$ 8        &   0.678 $\pm$ 0.022    &   0.032              &          0 \\       
    32& N  &     tuc &   0.3                &   0.932 $\pm$ 0.046        &   215 $\pm$ 8        &   0.791 $\pm$ 0.033    &   0.017              &          0 \\ \hline            
      & N  &     ALL &   0.3                &   0.760 $\pm$ 0.135        &   246.25 $\pm$ 29.54 &   0.751 $\pm$ 0.052    &   0.025  $\pm$ 0.006 &          0 \\ \hline        
   \end{tabular}
\end{table*}

\begin{table*}
  \caption{Initial conditions and results of our simulations with a
    cored DM profile: The
    first column is the number assigned to the simulation, second column
    is the dark matter profile, third column is the SFH,
    fourth column is the star formation efficiency, fifth, sixth and
    seventh column are the fitted parameters of the Sersic profile,
    eighth column is the clumpiness value and the last column is the
    number of surviving star clusters.}
  \label{tab:iniresp}
  \begin{tabular}{|l|l|l|l|l|l|l|l|l|}
    \hline
    Nr. & N/P & SFH & SFE & $\Sigma_{\rm eff}$ &  $R_{\rm eff}$ & n &
    C & Surv. \\         
      & & & & $\left[\frac{{\rm M}_{\odot}}{{\rm pc}^{2}}\right]$ &
    [pc] & & & SC(s) \\ \hline

33&  P  &      cte &   0.2               &   0.205 $\pm$ 0.013         &   469 $\pm$ 22         &   0.825 $\pm$ 0.050   &   0.107              &          2 \\  
34& P  &      cte &   0.2               &   0.189 $\pm$ 0.011        &   493 $\pm$ 21         &   0.933 $\pm$ 0.048   &   0.126                     &    2 \\       
35&  P  &      cte &   0.2               &   0.130 $\pm$ 0.007        &   593 $\pm$ 26        &   0.870 $\pm$ 0.044   &   0.112                &          0 \\      
36 & P  &      cte &   0.2               &   0.159 $\pm$ 0.005        &   533 $\pm$ 13         &   0.929 $\pm$ 0.027   &   0.082                  &      1 \\   \hline
  &  P &       ALL &   0.2               &  0.170 $\pm$ 0.033        &   522 $\pm$ 54.19        &  0.889 $\pm$ 0.051   &  0.106 $\pm$ 0.018      &      1.25 $\pm$ 0.957 \\ \hline
                                                                                37& P  &      lin &   0.2               &   0.202 $\pm$ 0.005        &   462 $\pm$ 8          &   0.903 $\pm$ 0.021   &   0.058                 &          0 \\      
38& P  &      lin &   0.2               &   0.200 $\pm$ 0.016        &   471 $\pm$ 27         &   0.939 $\pm$ 0.065   &   0.059                    &    1 \\             
39& P  &      lin &   0.2               &   0.127 $\pm$ 0.003        &   609 $\pm$ 10        &   0.855 $\pm$ 0.017   &   0.069                    &     0 \\        
40& P  &      lin &   0.2               &   0.159 $\pm$ 0.003        &   547 $\pm$ 9         &   0.824 $\pm$ 0.017   &   0.078                  &      0 \\  \hline 
  & P  &      ALL &   0.2               &   0.172 $\pm$ 0.035        &   522.25 $\pm$ 69.26      &   0.880 $\pm$ 0.050   &   0.066 $\pm$ 0.009                  &      0.25 $\pm$ 0.5 \\ \hline
                                                                               
41 & P  &      quad &   0.2              &   0.212 $\pm$ 0.005        &   451 $\pm$ 8         &   0.883 $\pm$ 0.020   &   0.038                  &      0 \\        
42 & P  &      quad &   0.2              &   0.206 $\pm$ 0.003        &   468 $\pm$ 5        &   0.818 $\pm$ 0.012  &   0.057                &          1 \\        
43 & P  &      quad &   0.2              &   0.130 $\pm$ 0.002        &   596 $\pm$ 9         &   0.869 $\pm$ 0.015   &   0.065                  &      0 \\        
44 & P  &      quad &   0.2              &   0.172 $\pm$ 0.003        &   520 $\pm$ 8        &   0.786 $\pm$ 0.016    &   0.054                  &      0 \\ \hline 
  & P  &      ALL  &   0.2              &   0.180 $\pm$ 0.037        &   508.75 $\pm$ 65.15       &  0.839 $\pm$ 0.045    &   0.053  $\pm$ 0.011                &      0.25 $\pm$ 0.5\\ \hline
                                                                                45 & P  &      tuc &   0.2               &   0.257 $\pm$ 0.010        &   408 $\pm$ 11         &   0.734 $\pm$ 0.030   &   0.060                &        1 \\            
46 & P  &      tuc &   0.2               &   0.234 $\pm$ 0.014        &   440 $\pm$ 18        &   0.736 $\pm$ 0.045   &   0.071                &        1  \\      
47& P  &      tuc &   0.2               &   0.149 $\pm$ 0.006        &   546 $\pm$ 16        &   0.789 $\pm$ 0.029   &   0.096                  &        0 \\       
48 & P  &      tuc &   0.2               &   0.151 $\pm$ 0.004       &   552 $\pm$ 10        &   0.924 $\pm$ 0.019   &   0.052                   &        0 \\ \hline
& P  &      ALL &   0.2               &   0.197 $\pm$ 0.055       &   486.5 $\pm$ 73.38       &  0.795 $\pm$ 0.089   &   0.069 $\pm$ 0.019                  &        0.5 $\pm$ 0.577\\ \hline
                                                                               
49&  P  &      cte &   0.3               &   0.336 $\pm$ 0.033         &   350 $\pm$ 26         &   0.504 $\pm$ 0.057   &   0.183                &      12  \\      
50 &  P  &      cte &   0.3               &   0.241 $\pm$ 0.022        &   446 $\pm$ 33        &   0.657 $\pm$ 0.054   &   0.186                  &      13  \\      
51&  P  &      cte &   0.3               &   0.206 $\pm$ 0.029         &   456 $\pm$ 50        &   0.562 $\pm$ 0.081   &   0.330                 &      9 \\        
52&  P  &     cte &   0.3              &   0.309 $\pm$ 0.029      &     359 $\pm$ 25        &   0.494 $\pm$ 0.053   &   0.154               &      6 \\ \hline   
&  P  &      ALL &  0.3                &  0.273 $\pm$ 0.059        &     402.75 $\pm$ 55.98   &  0.554 $\pm$ 0.074   &  0.213 $\pm$ 0.079                 &      10 $\pm$ 3.162 \\ \hline        
                                                                                 53&   P  &      lin &   0.3               &   0.299 $\pm$ 0.040        &   374 $\pm$ 39        &   0.600 $\pm$ 0.079   &   0.128                  &      9  \\       
54&  P  &      lin &   0.3               &   0.046 $\pm$ 0.026        &   1427 $\pm$ 72        &   1.894 $\pm$ 0.378   &   0.171                  &      11 \\       
55&   P  &      lin &   0.3               &   0.149 $\pm$ 0.014        &   538 $\pm$ 42        &   0.874 $\pm$ 0.056   &   0.128                  &      6  \\       
56&   P  &      lin &   0.3               &   0.193 $\pm$ 0.021        &   483 $\pm$ 43        &   0.807 $\pm$ 0.064   &   0.120                  &      5  \\ \hline
 &  P  &      ALL &   0.3               &   0.171 $\pm$ 0.104        &   705.5 $\pm$ 485.804        &   1.043 $\pm$ 0.578   &   0.136 $\pm$ 0.023   &      7.75 $\pm$ 2.753 \\ \hline    
                                                                                57&   P  &     quad &   0.3               &   0.288 $\pm$ 0.017        &   387 $\pm$ 18         &   0.641 $\pm$ 0.035   &   0.132                 &      9  \\       
58&   P  &      quad &   0.3              &   0.166 $\pm$ 0.026        &   539 $\pm$ 73         &   1.045 $\pm$ 0.099   &   0.119                 &      10 \\       
59&   P  &      quad &   0.3              &   0.160 $\pm$ 0.013        &   534 $\pm$ 37         &   0.720 $\pm$ 0.050   &   0.102                 &      5   \\      
60&  P  &      quad &   0.3              &   0.185 $\pm$ 0.016        &   501 $\pm$ 36         &   0.780 $\pm$ 0.051   &   0.089                 &      5  \\ \hline
  &   P  &      ALL  &   0.3              &   0.199 $\pm$ 0.059        &   490.25 $\pm$ 70.86  &   0.796 $\pm$ 0.175   &   0.110 $\pm$ 0.018              &      7.25 $\pm$ 2.629 \\ \hline
                                                                                61&  P  &      tuc &   0.3               &   0.341 $\pm$ 0.026        &   346 $\pm$ 19        &   0.526 $\pm$ 0.043   &   0.122                  &      9   \\      
62&  P  &      tuc &   0.3               &   0.221 $\pm$ 0.039        &   440 $\pm$ 64        &   0.847 $\pm$ 0.108   &   0.136                  &      9 \\        
63&  P  &      tuc &   0.3               &   0.183 $\pm$ 0.029        &   483 $\pm$ 61        &   0.677 $\pm$ 0.092   &   0.293                  &      7  \\       
64&  P  &      tuc &   0.3               &   0.064 $\pm$ 0.009        &   927 $\pm$ 114       &   1.728 $\pm$ 0.096   &   0.179                  &      5  \\ \hline
  &  P  &      ALL &   0.3               &   0.202 $\pm$ 0.114        &   549 $\pm$ 258.41    &   0.944 $\pm$ 0.538   &   0.182 $\pm$ 0.077      &      7.5 $\pm$ 1.914 \\ \hline
   \end{tabular}
\end{table*}

\begin{figure*}
  \includegraphics[width=130mm]{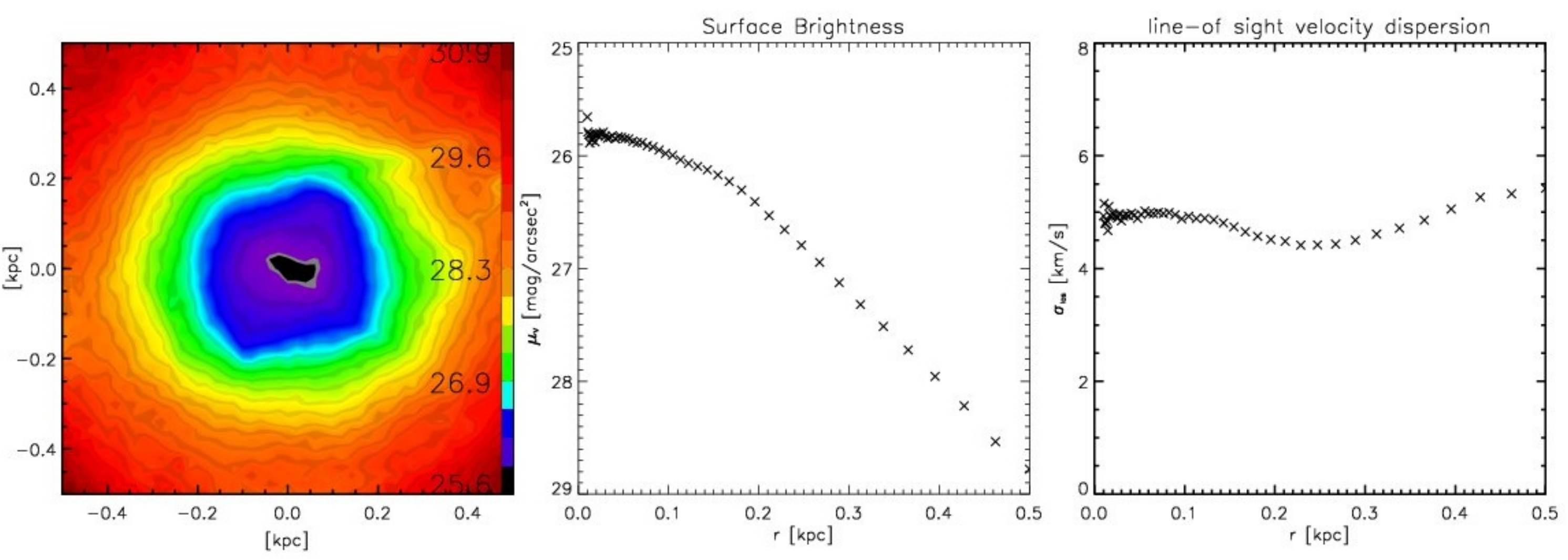}
  \includegraphics[width=130mm]{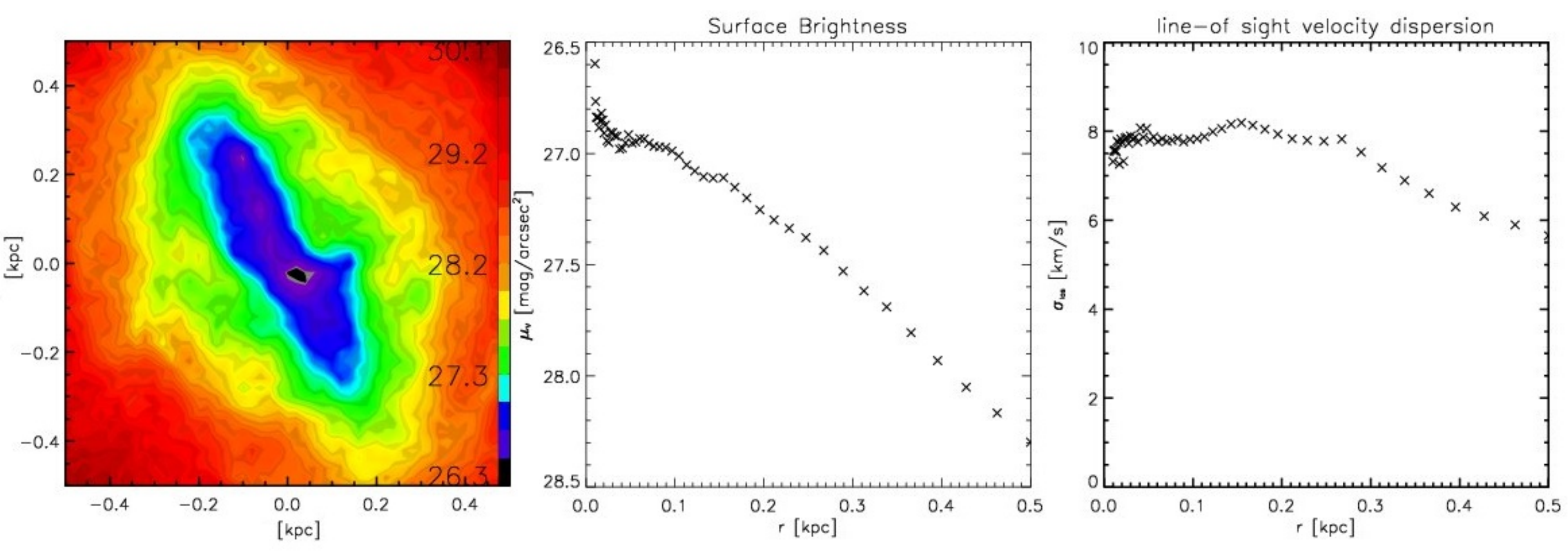}
  \includegraphics[width=130mm]{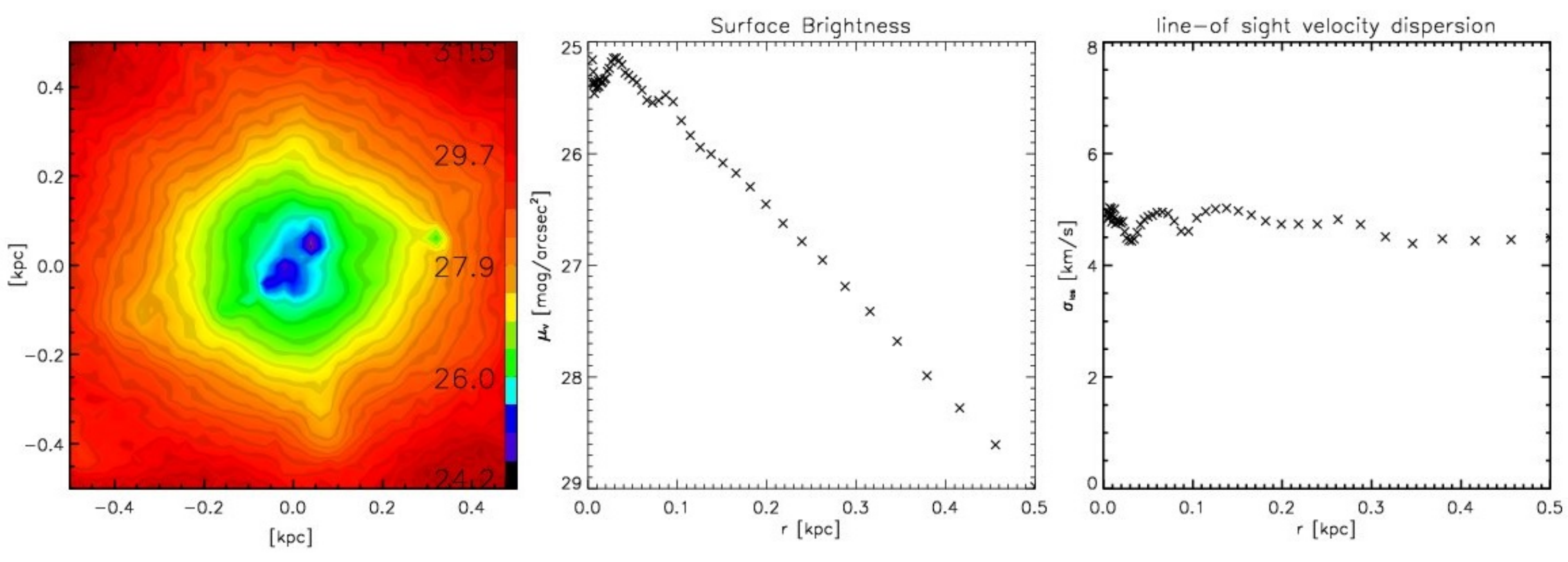}
  \includegraphics[width=130mm]{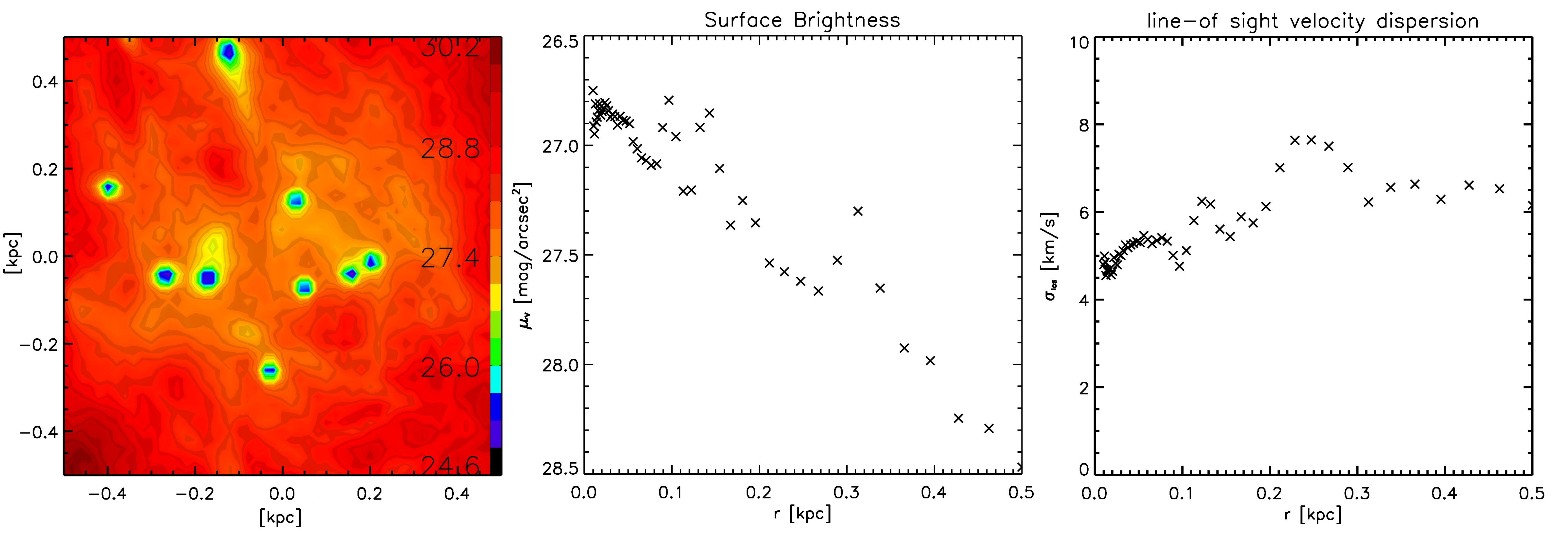}
  \caption{Characteristics of the final object inside a radius of
    500~pc, the left panel is the shape of the final object with 
    color bars for the magnitudes of surface brightness, the centre
    panel shows the radial surface brightness profile and the right
    panel is the radial velocity dispersion profile, both measured in
    concentric rings.  First row shows a simulation with a a linear SFH,
    SFE=20\% and NFW halo (simulation 5), second row: quadratic SFH,
    SFE=20\% and cored DM halo (simulation 41), third row: constant SFH,
    SFE=30\% and NFW halo (simulation 20), last row: constant SFH,
    SFE=30\% and cored DM halo (simulation 50). The simulation numbers are
    according to the list in Tab.~\ref{tab:iniresn} \& \ref{tab:iniresp}.}
  \label{fig:res}
\end{figure*}

We consider the following setup for our simulations:

\begin{enumerate}[i]
\item  Simulations with a cusped dark matter halo, as cosmological
  simulations predict, use a Navarro, Frenk \& White
  \citep[NFW;][]{Navarro1997} profile, using 1,000,000 particles 
  according to the recipe described in \citet{Dehnen2005}.  We use a
  scalelenght for the halo of $R_{\rm s,h} = 1$~kpc and define an
  enclosed halo mass within 500~pc of $M_{500} = 10^7$~M$_{\odot}$,
  following the fiducial model described in \citet{Assmann2013a}.
  Using a standard value for $H_0 = 70$km\,s$^{-1}$Mpc$^{-1}$ and
  $r_{\rm vir} = r_{200}$, we obtain a virial radius of $r_{\rm vir} =
  12.7$~kpc, i.e.\ the concentration of the halo is $c = 12.7$, which
  agrees with the range of values found with dSph galaxies of
  5-20.  The total mass of the halo out to the virial radius amounts
  to $2.3 \times 10^8$~M$_{\odot}$.  

\item  Simulations with a cored dark matter halo use a
  Plummer sphere profile \citep{Plummer1911}, using 1,000,000 particles,
  with a scale-length (Plummer radius) of $R_{\rm s,h} = 1$~kpc, and an
  enclosed mass within 500~pc of $M_{500} = 10^7$~M$_{\odot}$.  For
  the Plummer profile we use a cut-off radius of $R_{\rm cut} =
  5$~kpc, which contains more than 94\% of the total mass of a Plummer
  sphere, accounting in our case to a total mass of the halo of $1.1
  \times 10^8$~M$_{\odot}$.  

\item  For the luminous component, the star clusters, we use $N = 30$
  star clusters.  This number is chosen arbitrarily, because the
  previous results of \citet{Assmann2013a,Assmann2013b}, show no
  significant differences using different numbers of SCs.  As it is
  claimed that dSph galaxies had low star formation rates and
  therefore also a low SFE \citep{Bressert2010}, we form rather low
  mass open clusters and associations than massive SCs.  Each SC is
  modelled as a Plummer sphere \citep{Plummer1911}, with 100,000
  particles using the recipe of \citet{Aarseth1974}.  The SCs have a
  Plummer radius (half light radius) of $R_{\rm pl} = 4$pc and a cut-off
  radius of 25pc.  This is similar to the radii found for young SCs in
  the Antennae \citep{Withmore1999}.  

\item  We mimic the gas-expulsion of the SCs by artificially reducing
  the mass of each particle until the final mass is reached after one
  crossing time, i.e.\ 4~Myr.  As the mass in lost gas is negligible
  compared to the DM mass we do not take this mass further into
  account.  The final mass in stars (of all SCs) after this mass loss
  amounts to $4.5 \times 10^5$M$_{\odot}$, which is a typical stellar
  mass of  one of the classical dSph (e.g. \citet{Mateo1998}).  We
  perform simulations with SFEs of 20\% and 30\%.  Therefore, the
  initial mass of each SC in its embedded phase is $5 \times
  10^4$~M$_{\odot}$ for SFE = 30\% and $7.5\times 10^4$M$_{\odot}$ for 
  SFE = 20\%.  Note, that this results in a particle resolution in our
  simulation which is slightly better than reality, i.e.\ more star
  particles than actual stars (see explanation below).  

\item  The star clusters themselves are distributed in virial
  equilibrium inside the halo according to a Plummer distribution.
  This is made for simplicity, because we do not know exactly in which
  virial state the SCs form with respect to the halo and we expect
  more star clusters to form in the centre than in the outer parts.  
  The distribution has a scale-length of $R_{\rm pl,sc} = 0.25$~kpc
  and a cut-off radius of $R_{\rm cut-off} = 1.125$~kpc.  The initial
  orbital velocities are obtained from the Jeans equation:
  \begin{eqnarray}
    \label{eq:jeans}
    \sigma_{r,i}^2(r)& = & \frac{1}{\rho_{i}(r)} \int_{r}^{rc}
    \frac{GM_{tot}(r')}{r^2}\rho_{i}(r')dr' ,
  \end{eqnarray}
  where $M_{\rm tot}$ correspond to the total mass given by the sum of
  the mass of all SCs and the mass of the DM halo.  We do not assume
  any anisotropy in spatial nor in the velocity distribution.  Also we
  do not take into account that the SCs might form in a disc-like
  structure showing angular momentum in their distribution, as we
  assume that the gas distribution on the small scales of a dSph is
  rather supported by pressure than rotation.  With this assumption we
  differ from most of the previous models in which dSph galaxies are
  simply the evolutionary outcome of harassed dwarf disc galaxies,
  which lost their angular momentum because of gravitational forces.
  
\item  To mimic the star formation histories we insert the star
  clusters at different times into the simulations.  We probe four
  different star formation histories.  Fig.~\ref{fig:sfh} shows the
  SFHs used in this project.  The first is a constant SFH in which we
  insert the star clusters one by one over the whole 10~Gyr of
  evolution in equally spaced time intervalls.  The second SFH mimics
  a linearly declining SFR and the third a quadratic decline in SFR.
  I.e.\ in both cases the majority of SCs are inserted at early times
  of the simulation and only a few get inserted at late times.  The
  last SFH mimics a double burst as implied for Tucana.  Half of the
  SCs get inserted at the beginning of the simulation and the second
  half after 5~Gyr, i.e.\ half of the simulation time.  For the common
  SFH of dSph galaxies, the single burst at very early times, we refer
  to the models of \citet{Assmann2013a,Assmann2013b} in which the
  simulations start with all SCs from the beginning. 
\end{enumerate}

\begin{figure*}
  \includegraphics[width=180mm]{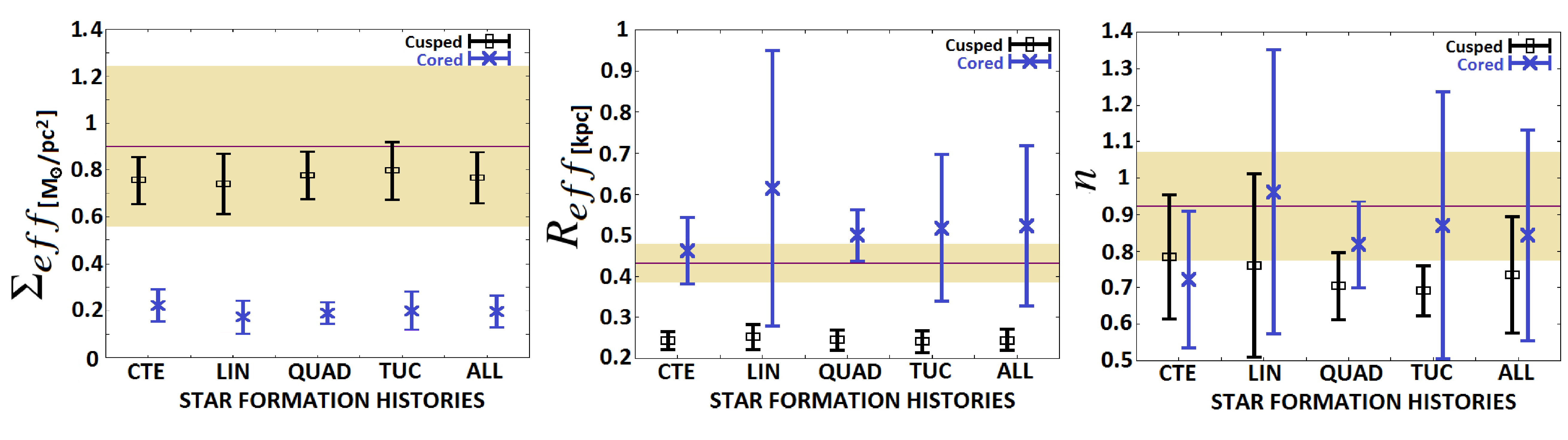}
  \caption{Mean values of the Sersic profile fits for different DM
    halos and SFHs.  The data points are mean values from all
    simulations using the same DM halo and both values of SFE (0.2 and
    0.3).  Previous and our results have shown that the profile
    parameters do not depend on the SFE used.  As can be seen in the
    panels, it also does not depend on the SFH used.  Therefore, we
    plot as the last point the mean value for all simulations with the
    same DM halo, independent of SFE and SFH.  Black bars are for
    cusped DM halos and blue crosses are for cored DM
    halos, the purple line shows the results of the previous work
    \citep{Assmann2013b}.  The left panel shows the effective surface
    brightness, the middle panel the effective radius and the right panel
    the Sersic index $n$.} 
  \label{fig:sersic}
\end{figure*}

We simulate the cluster complex within the DM halo using the particle
mesh code {\sc Superbox} \citep{Fellhauer2000}.  {\sc Superbox} has
two levels of high resolution sub-grids.  The highest resolution grid
has a chosen resolution (i.e.\ cell-length) of 8.3~pc for the dark
matter halo and 0.41~pc for the star clusters and covers the central
area of the halo or the SCs completely, respectively.  The medium
resolution grid has a cell-length of 41~pc for the DM halo and 4.1~pc
for the SCs.  Finally, the outermost grid covers the complete area
beyond the virial radius of the dark matter halo with a resolution of
830 pc.  The time-step is fixed at 0.25~Myr to resolve the internal
dynamics of the SCs and we simulate for 10~Gyr.
 
As the SCs dissolve immediately, two-body relaxation effects are not
important, i.e.\ we are able to use a fast particle-mesh code.  A
particle mesh-code naturally neglects close encounters between
particles (which here are rather representations of the phase space
than actual single stars) and is therefore called collision-less.  That
the particles are phase-space representations makes it (in our case)
possible to actually use more particles than actual stars.  With the
same reasoning we can model the DM halo without using an actual number
of DM particles. 

We run 64 simulations for 10~Gyr.  For the setup of the simulations we
refer to the first columns in Tabs.~\ref{tab:iniresn} \& \ref{tab:iniresp}. 

\section{Results}
\label{sec:results}

In Fig.~\ref{fig:res}, we show a set of four different simulations as
examples.  The examples show a wide range of different outcomes from our
simulations, but all except the last one show properties similar to the
classical dSph galaxies of the MW.

Our models predict objects with different shapes and in
Fig.~\ref{fig:res} we show: a) a final object nearly spherical in
which all the SCs have dissolved (simulation number 5), b) a very elongated
final object in which all SCs have dissolved, but some of them had similar
orbital planes causing a flattened profile (number 41), c) a double cored
final object with two high density peaks in the centre (number 20) and
finally d) a final object which can not resemble the properties of
classical dSph because the SCs did not dissolve completely (number 50).
The number point to the respective simulations in Tabs.~\ref{tab:iniresn}
\& \ref{tab:iniresp}.

In the next subsections we will focus on the analysis of the surface
brightness profile, the velocity dispersion, the shape of the final
object and the surviving SCs.  A summary of important results can be
found in Tabs.~\ref{tab:iniresn} \& \ref{tab:iniresp}.

\subsection{Surface brightness profile}
\label{sec:surf}

We fit the surface brightness profile of our simulations using a
Sersic profile:
\begin{eqnarray}
  \label{eq:sersicfit}
  \Sigma (R) & = & \Sigma_{\rm eff} \exp \left( -b_{\rm n}
    \left[\left(\frac{R}{R_{\rm eff}}\right)^\frac{1}{n}-1\right]\right)
  \\ 
  b_{\rm n} & = & 1.9992n-0.3271 \nonumber
\end{eqnarray}
with $R_{\rm eff}$ the effective radius and $\Sigma_{\rm eff}$
the surface density at the effective radius.  This profile has the
advantage of having three free parameters, so our simulations can be
fitted easily.  The index $n$ gives us information about the shape of
the dSph galaxy.  In the case $n \approx 1$, we will have an
exponential profile for the surface density distribution, as it is
observed in classical dSph galaxies
\citep{Caon1993,Jerjen2000,Walcher2003}.  

\begin{figure*}
  \includegraphics[width=160mm]{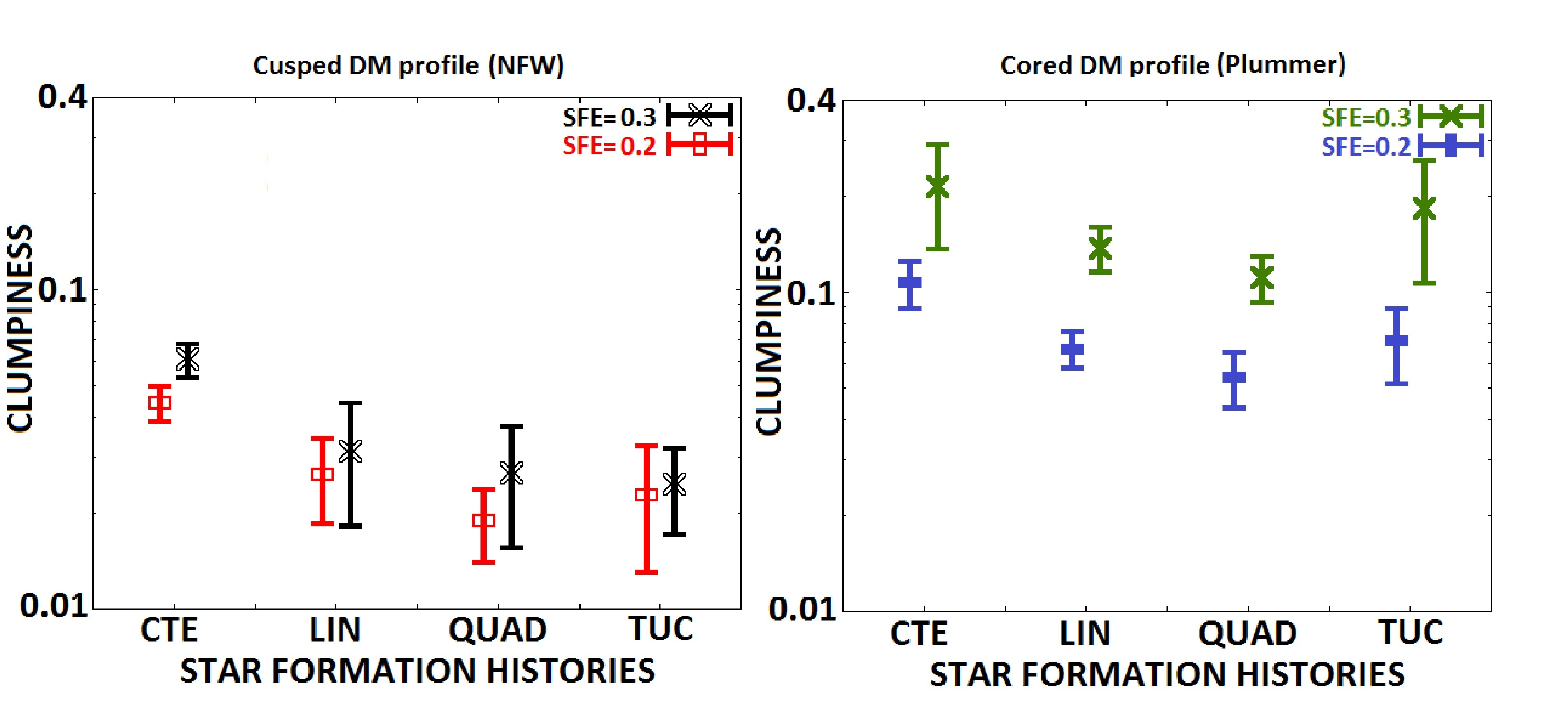}
  \caption{Mean clumpiness values for different SFHs and SFEs in a
    cusped DM halo (left panel) and a cored DM halo (right panel).
    Simulations with SFE=30\% are shown in black and green crosses
    and SFE=20\% are denoted by red and blue bars.  Note, that the
    $y$-axis has a logarithmic
    scale.  Cored DM halo models have higher clumpiness values than
    cusped models, furthermore recent star formation and higher SFE
    give us more inhomogeneous final objects.}  
  \label{fig:clump}
\end{figure*}

Surface brightness profiles show integrated light along a line-of-sight.
As we do not know the actual line of sight, towards our
theoretical objects, we project our objects along the three different 
Cartesian coordinates.  Then we bin the data in 50 concentric rings
with logarithmically equally spaced widths from the centre out to
$0.5$~kpc.  We calculate a mean value profile with one sigma deviations
from the three 
projections.  That way we avoid to take the elliptic shape of the
contours into account and we do not need to project along the major
axes of the object.  This is the same procedure to produce the data as 
described in \citet{Assmann2013a,Assmann2013b}, i.e.\ it is possible
to compare our data directly with the previous results.  Furthermore,
as stated above, we do not know the exact orientations the classical
MW dSph are presenting towards us.  As long as those orientations are
unclear, it makes more sense to compare our mean model profiles to the
randomly oriented real dSph galaxies, than to compute values along the
main axes of our objects.

In Fig.~\ref{fig:sersic}, we show the correlation between the three
parameters of the Sersic profile ($\Sigma_{\rm eff}$, $n$ and $R_{\rm
  eff}$), fitted to our models, versus their SFH for the different DM
halos.  Our results do not show any dependence of the SFE used,
thereby confirming the results of \citet{Assmann2013a,Assmann2013b},
and the mean values shown are already averages over the different SFEs
used.  We also see that there is no dependency on the SFH in our data.
Therefore, the final data point in each panel shows a mean value of
all simulations using the same DM halo.  The purple lines are the mean
values of the previous study by \citet{Assmann2013a,Assmann2013b}.  The
one sigma deviations are shown as shaded areas around the mean.
There is indeed one word of caution left.  In the presentation of
their fiducial model in \citet{Assmann2013a}, the mean values quoted
there are similar to the cored profile results in our study.  Only if
we use all simulations with the same parameters ($M_{500}$, $R_{\rm h}$ and
$R_{\rm sc}$) from \citet{Assmann2013b} we obtain similar results as in
our study.  This shows that the random placement of a small $N$ number of
SCs can lead to
a huge variation in the final results, even when using the same over-all
parameters.  As an example we quote the minimum and maximum values for
this parameter set from \citet[][see results given
  in the tables]{Assmann2013b}: for $\Sigma_{\rm eff}$ we obtain
values from $0.1$ to $3$~M$_{\odot}$\,pc$^{-2}$, for $n$ from $0.4$ to $2.1$
and for $R_{\rm eff}$ from $140$ to $740$~pc.

Taking this words of caution into account, we still see that the mean
effective surface brightness is higher for NFW halos than for the cored
Plummer halos.  The opposite is true for the effective radius; cored
profiles show about twice the effective radius than cusped profiles.
Here we are in disagreement with \citet{Assmann2013b}, which see no
significant differences between the two types of profiles.  The answer
to this riddle is that for our study we have more simulations per
parameter set (but a smaller range of parameters) than used in
\citet{Assmann2013b}.

Fig.~\ref{fig:sersic} shows clearly that cusped DM halos give us higher
values for $\Sigma_{\rm eff}$, and lower for $R_{\rm eff}$ than cored
DM models.  This is because a cusped profile has a high density in the
centre and the stars are more likely to get stripped in the central
part of the DM halo than in a cored DM profile. 

We see that the Sersic index $n$ has no dependency on any initial
parameter.  It shows always mean values around 1, independently of the
SFH or DM halo.  This means that our resulting objects have
approximately an exponential surface brightness distribution, like the
dSph galaxies of the Local Group
\citep{Walcher2003,Jerjen2000,Caon1993}. 

Taking all simulations into account we obtain the following mean
values: For our NFW halo models a Sersic index of $n = 0.74 \pm 0.16$,
an effective radius of $R_{\rm eff} = 0.245 \pm 0.026$~kpc and a
surface density at this radius of $\Sigma_{\rm eff} = 0.77 \pm
0.11$~M$_{\odot}$\,pc$^{-2}$.  For the cored Plummer profiles as DM
halo we measure $n = 0.84 \pm 0.29$, $R_{\rm eff} = 0.52 \pm 0.20$~kpc
and $\Sigma_{\rm eff} = 0.196 \pm 0.068$~M$_{\odot}$\,pc$^{-2}$. 

Photometrical observations of the dSph galaxies in the Local Group
show that we can find different sizes of these galaxies.  For example,
the effective radii of Draco, UMi, Sculptor and Fornax are 180, 200,
110 and 460 pc, respectively \citep{Mateo1998,Irwin1995}, i.e.\ span
the complete range between our cusped and cored models. 

Furthermore, in this study we kept the radius of the SC distribution
constant in all simulations.  As stated in \citet{Assmann2013b} our
models predict an effective radius of the luminous component of
about/up to twice the size of the star cluster distribution.  Our new
results show that this is still true for cored DM halo profiles but
are closer to equal in size in the case of NFW halos.

\subsection{Clumpiness}
\label{sec:clumpiness}

We now consider the clumpiness parameter to characterise the shape of
our resulting object.  The clumpiness is a way to measure the
inhomogeneity in the distribution of stars.  We use the method
explained in \citet{Conselice2003} and construct a smooth elliptic
model which fits our simulation result, using the IRAF routine
ELLIPSE.  Then, we subtract this smooth model from our data and sum
the positive residuals.  The ratio between these residuals and the
original data is the clumpiness C: 
\begin{eqnarray}
  \label{eq:clump}
  C & = & \frac{\Sigma_{\rm allpix} m_{\rm residual,pixel}}
  {\Sigma_{\rm allpix} m_{\rm original,pixel}}. 
\end{eqnarray}

In Fig.~\ref{fig:clump}, we show the relationship between the
clumpiness of our models the SFH, SFE and DM distribution.  The left
panel is for cusped DM halo models and the right panel is for cored DM
haloes.  Crosses are for SFE=30\% and bars for SFE=20\%
and in the x-axis we plot the SFH. 

First, we observe that all our simulations lead to low values of the
clumpiness parameter.  This is a hint that the substructure we see in
the brightness maps of our model might be only visible due to our
larger than reality particle resolution.  But those faint structures
in our simulations are definitely real and not due to noise and the
brighter ones should be observable.   

These plots show that models with recent star formations have higher
values of clumpiness than models without.  The later the star clusters
form the less time they have to dissolve and form a homogeneous
final object.  Furthermore, the value of clumpiness depends on the
SFE: If the SFE is higher (30\%) the star clusters will need more time
to dissolve after gas-expulsion leading to higher values of clumpiness
than models with low SFE (20\%). 

We observe that in a cored DM profile we measure higher clumpiness
parameters than in cusped models.  This is a consequence that Plummer
models have central crossing times larger than models with NFW profile
and less interaction time to erase substructure.  Also a cusp is more
efficient in dispersing stars from their original orbits, leading to a
more homogeneous distribution with time.

The large error bars in the clumpiness value for cored DM halos and
SFE=30\% are due to the high number of surviving SCs, so our final
object is not easy to fit.  

It is hard to compare our results with the classical dSph of the MW as
there is no determination of their clumpiness published, but we
observe in classical dSph the same elongations, twists, deviations from
ellipses and/or double cores \citep[see e.g.][]{Irwin1995} that elevate
the clumpiness values in our models. 

\begin{figure*}
  \includegraphics[width=150mm]{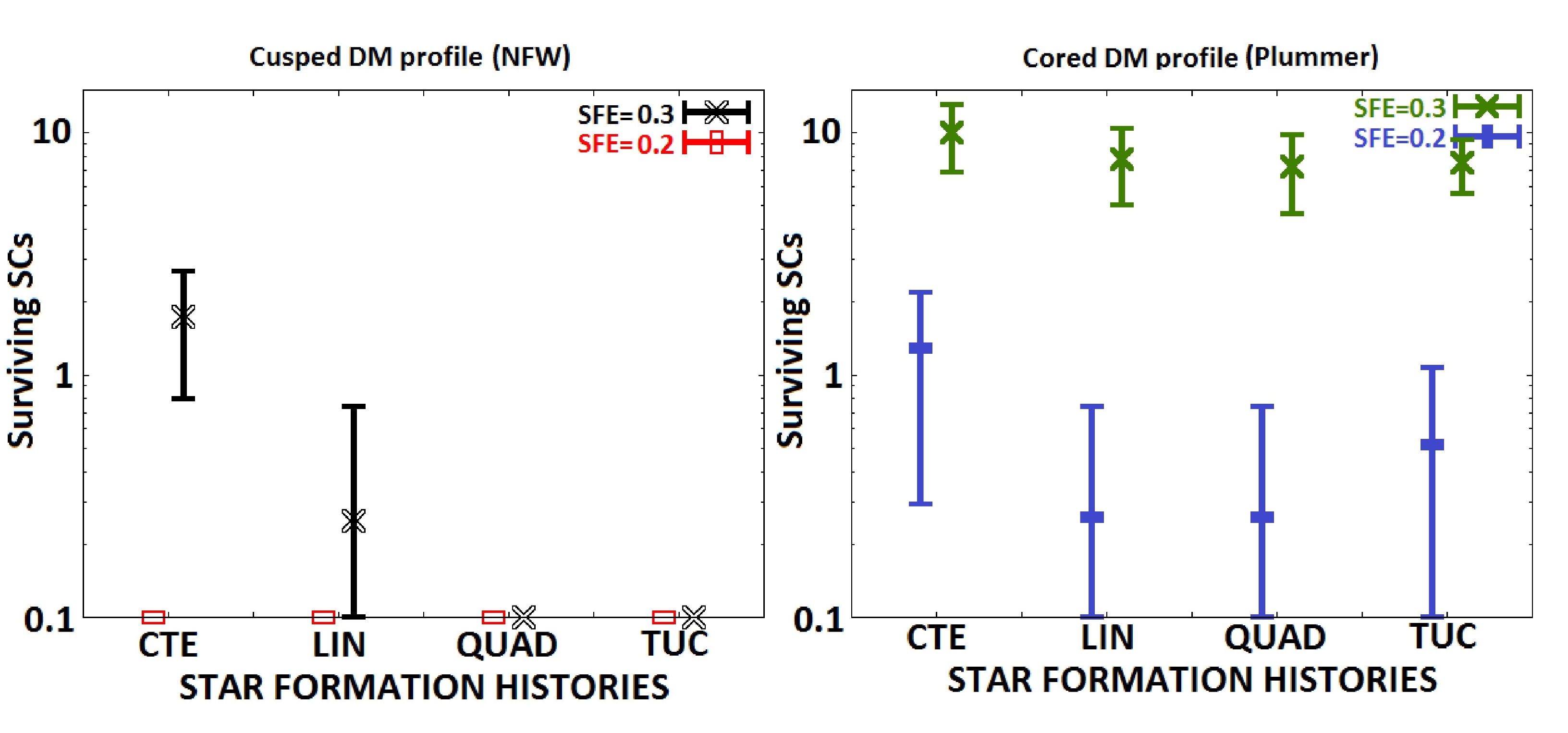}
  \caption{Number of surviving SCs vs.\ SFH; the left panel shows
    simulations with a cusped DM halo profile and in the right panel are the
    simulations with a cored DM profile.  Red and blue bars denote
    SFE=20\% and black and green crosses SFE=30\%.  The $y$-axis has a
    logarithmic scale and therefore 0 surviving SCs are marked as 0.1 in
    the plots.  In cusped DM halos the SCs dissolve easily and we do not
    see any survivors in simulations with no recent star formation.
    In cored DM halos we have several surviving star clusters even with
    a SFE=20\% and no recent star formation.}  
  \label{fig:ssc}
\end{figure*}

\subsection{SC dissolution}
\label{sec:dissolve}

Some of our final objects have several surviving SCs which are
orbiting the centre of the DM halo.  Those SCs can survive either
because they need more time to get dissolved or because their orbit is
very close to the centre of the DM halo where the stars struggle to
escape the deep potential well of the halo.  Also cored DM halos need
more time to dissolve SCs than cusped DM halos as shown also in e.g.\
\citet{pen09} and \citet{amorisco2017}.  We consider a SC to be dissolved
if it has less than 10\% of particles bound.  In Fig.~\ref{fig:ssc} we
show the correlation between the number of surviving SCs, the SFH, SFE
and the DM profile. 

In simulations with low SFE (20\%) we see no surviving
SCs in cusped DM halos, because it only takes them a few hundred Myr
to dissolve.  In simulations with cored DM haloes, the dissolution time
of SCs is significantly longer.  Therefore, especially in simulations
with late star formation (constant SFH) most of the SCs, inserted
last, have no time to dissolve. 

\begin{figure}
  \includegraphics[width=86mm]{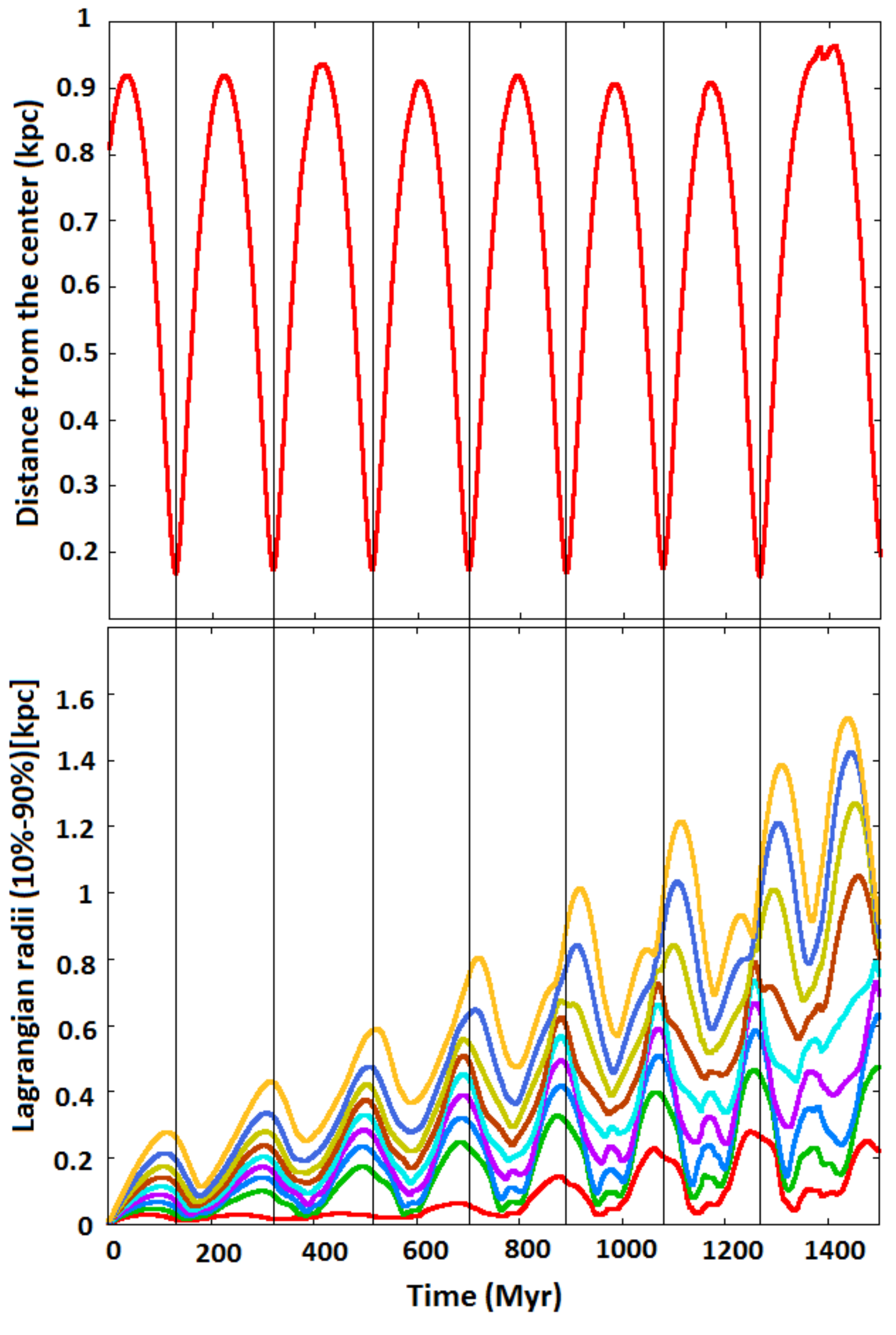}
  \caption{This plots are taken from a SC in a cored DM halo and a
    SFE=30\%.  The top panel shows orbit of the SC, i.e.\ the distance
    to the centre of the DM halo, and the bottom panel the Lagrangian
    radii (10\%-90\%), measured for the original SC particles with respect
    to the present SC location.  We see that the SC is
    expanding and contracting while it is orbiting the DM halo.  The
    SC expands while it is passing through peri-centre and contracts
    at its apo-centre.} 
  \label{fig:rad}
\end{figure}

In simulations with high SFE (30\%), the number of surviving SCs is
higher.  For cusped DM profiles we need approximately $1$~Gyr to
dissolve the SCs.  For cored DM profiles the number of surviving SCs
is much higher, and there are some star clusters which cannot be
dissolved even after $10$~Gyr if the apocentric distance is small ($<
300$~pc).  In Fig.~\ref{fig:rad} we show the process of dissolution of
a SC orbiting the centre of a DM halo.  We see that the SC is
expanding while it is orbiting the centre of the DM halo.  It expands
when near the centre and contracts again while far from the centre.
In that sense we do see sometimes fluffy density enhancements, which
could be mistaken for an extended cluster.  In reality we see a dissolved
SC close to its apocentre, where the original stars of the SC are
in compressed tidal tails, mimicking a density enhancement at its current
position.  This effect is more pronounced, if clusters are orbiting
closer to the centre of the DM halo.

Intuitively, we have more surviving SC in simulations with recent star
formation because the last SCs which are placed within the DM halo
have less time to get dissolved completely in comparison to the older
SCs. 

\begin{figure*}
  \includegraphics[width=150mm]{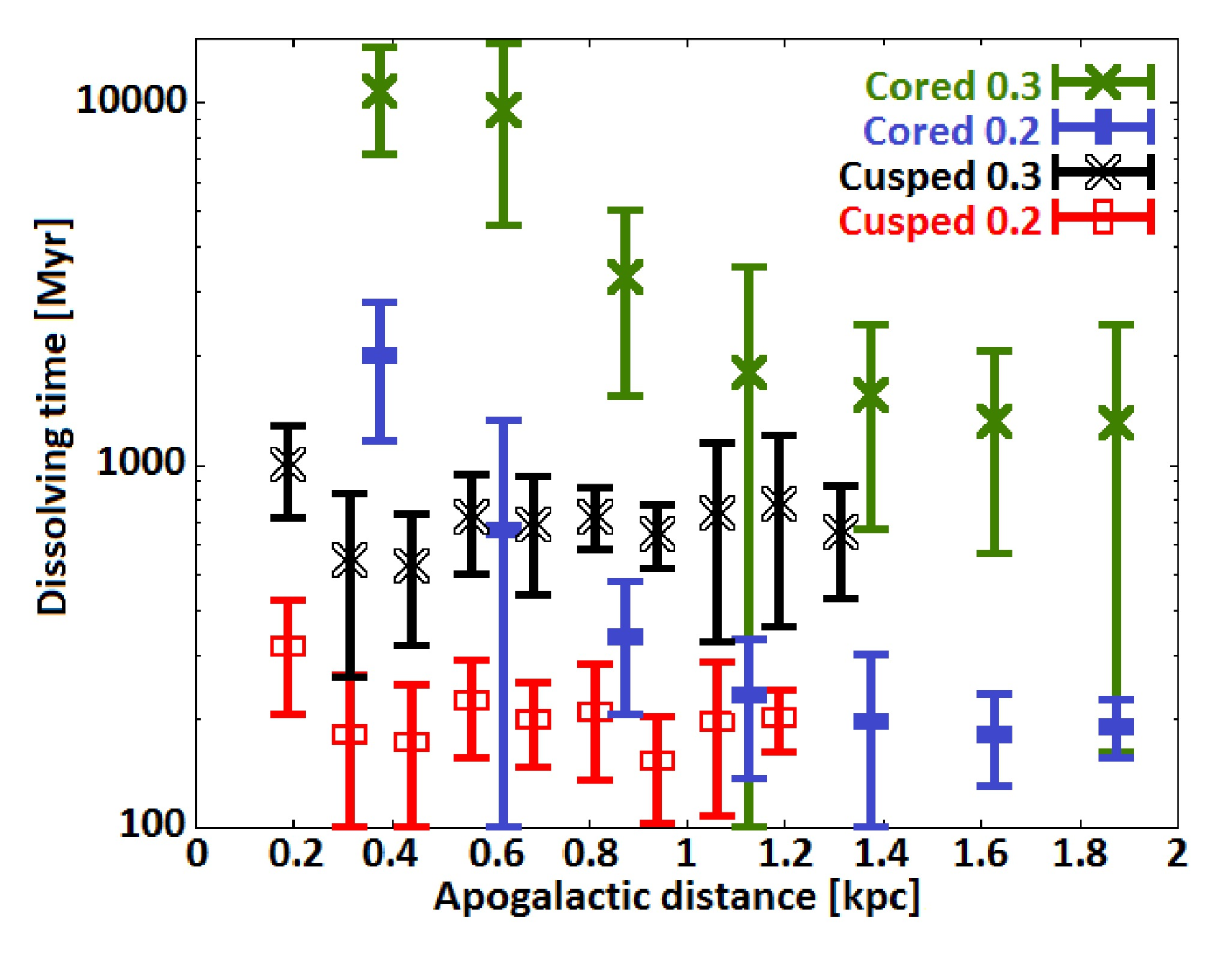}
  \caption{Dissolution time vs.\ apo-centric distance: This plot shows
    the dissolution time of the star clusters according to their orbit,
    dark matter profile of the halo and SFE.  Filled bars and crosses are
    for cored DM halos and a SFE of 0.2 or 0.3 respectively, open bars
    and crosses are for cusped DM halos with a SFE of 0.2 and 0.3
    respectively.  We see that the dissolution time is always longer for
    cored than for cusped halos and longer for higher SFEs.  Furthermore,
    we see a clear trend in the cored simulations that SCs orbiting closer
    to the centre (small apo-centres) have significantly larger
    dissolution times.}  
  \label{fig:diss}
\end{figure*}

The time to erase a SC will vary depending on its orbit, SFE and the
DM profile, in Fig.~\ref{fig:diss} we plot the correlation between the
dissolving time and the apocentric distance for both DM halos and
SFEs.  Filled bars and crosses are for the cored DM haloes (SFE=20\% and
30\% respectively), while open bars and crosses denote cusped haloes.
In order to say that a SC is dissolved we measure the time when less
than 10\% of the particles are bound.  

Fig.~\ref{fig:diss} shows that for a cusped DM profile the dissolving
time is always lower than in the corresponding cored simulations and
will not take more than 1~Gyr, even if the SFE is high (30\%).
Furthermore, we see that in cusped simulations the dissolution time is
almost independent from the apo-centric distance of the SCs.  On the
other hand for cored DM profiles the dissolution time is higher if the
apo-centric distance of the orbit is small ($< 300$~pc), this is 
because the stars of the SC can not escape easily from the potential
well of the core, while at the same time there is no steep potential
difference within the core.  So the stars get less dispersed and get
compressed along the orbit and re-captured close to the apo-centre.
SCs with large apo-centric distances (outside the central area) in
cored simulations will indeed suffer from strong potential gradients
along their orbit and therefore are dissolving faster.  If we increase
the SFE to 30\%, the SCs with the smallest orbits will need more than
10~Gyr to get dissolved.  

The presence of surviving SCs in our simulations, could give us
predictions for future observations to corroborate our formation
scenario, for example in Fornax there are five old SCs discovered
orbiting this galaxy and in several dwarf galaxies we see young star
clusters \citep[e.g.][]{grasha2017}. 

\subsection{Velocity space}
\label{sec:vel}

The dSph satellite galaxies of the MW are close enough to obtain high
resolution measurements of their line-of-sight velocity dispersions,
which are in the order of 5 to 10~km\,s$^{-1}$ \citep{Walker2007}.
We compare the line of sight velocity dispersion profile of the final
objects with the typical velocity dispersion observed in classical
dSph galaxies. 

To measure the line-of-sight velocity dispersion of our models we
consider the mean values of the velocity dispersion of all pixels
within a radius of 10~pc and 500~pc from the centre of the object.
This mean value is calculated by considering the mean velocity
dispersion along all three coordinate axis because the orientation of
our objects is unknown.  With this method we also keep the same
analysis of the data as in our previous studies to be able to compare
our results. 

In the right panels of Fig.~\ref{fig:res}, we show the line-of-sight
velocity dispersion profiles for some of our simulations.  These
profiles have been obtained considering the mean value of the line of
sight velocity dispersion of pixels within concentric rings with
varying radius.  In all simulations the velocity dispersion profiles
are always more or less flat, except in models with a high quantity of
surviving SCs like Fig.~\ref{fig:res} (lowest panels), all of our
models show velocity dispersion in the range of 5 to 12~km\,s$^{-1}$,
which is in agreement with observations. 

Looking at the outer part of the profiles we observe different types
of behaviour in the velocity dispersion.  We get outer profiles (beyond
1~kpc) where the velocity dispersion falls slowly in some cases, while
in others the dispersion stays flat.  We see a similar behaviour in the
MW's dwarf galaxies.  In Sextans we see a slight drop in velocity
dispersion around 1~kpc, while Sculptor, Draco and Fornax show flat
profiles \citep[][their figure~2]{Walker2007}.

In the inner part, some of our simulations show wiggles and bumps in
the dispersion profile.  While the observers always try to fit smooth
curves, thus implicitly assuming that any bumps seen in the profiles
are merely due to statistical noise (e.g.\ Sculptor, Draco or Carina),
in our models, the strange bumps in the observed profiles are not due
to noise in the observed data sets.  According to our formation theory
they are a natural product of the formation scenario proposed.  For a
detailed investigation of this phenomena we refer to our previous
publications \citep{Assmann2013a,Assmann2013b}.

Finally, in the central part we see the same behaviour as shown in the
multitude of MW's dwarf galaxies.  Some of our models have a central
dip in the velocity dispersion profile like in Sextans or Draco.
Others show a rising central velocity dispersion.  We do not see a
similar behaviour with the classical dwarf spheroidal of the MW but
there are hints that some of the faint dwarfs galaxies have an
elevated central velocity dispersion. 

These results show clearly that our models are well suited to
reproduce the formation of dSph galaxies.  We can show that we can
reproduce the dynamics of the different dwarf galaxies with our
models. 

\section{Discussion}
\label{sec:disc}

Our simulations can resemble the observed properties of classical
dSph, as for example an exponential surface brightness profile and a
flat velocity dispersion in the range of 5 to 12~km\,s$^{-1}$.
Furthermore, we can resemble specific properties seen in some dSph, as
for example off-centre nuclei (Sextans), secondary density peaks (Ursa
Minor) or dents in the contours (Draco).  

An important feature of our simulations is the presence or absence of
surviving SCs.  We identify three parameters which can be responsible
for the survival of a star cluster.  The most obvious one is a high
SFE, which automatically leads to survival of the gas-expulsion phase.
In our simulations, the SFEs used are below the threshold, which is
usually believed to produce a surviving star cluster, even though 30\%
is close to the theoretical limit.  Our simulations definitely show
that lower SFEs lead to less or none surviving SCs.  The next best
reason should be that in the case of more recent SFH some SCs have not
enough time to get completely dissolved.  Again, we see this trend in
our simulations.  But, we identify a third possibility for SCs to
survive and that is in the very centre of a cored DM halo.  A central
core does not disperse the stars of the SC from their original orbit
and so we see SC remnants, especially if the SC is close to its
apo-centric distance.  This might be a hint how to support the
formation of nuclei in dwarf galaxies.

Our simulations provide a possible solution to why some dSph galaxies
have orbiting star clusters and why some of them do not, and this
could be a hint to looking for faint surviving SCs in other dSph with
 future telescopes.  As the majority of dSph galaxies do not show
associated SC, with the exception of Fornax and Sagittarius, we can
deduce, that the SCs inside the DM halos have formed with low SFEs
($<30$\%) and most likely in a cusped DM profile as predicted by
$\Lambda$-CDM.  

The absence of surviving SCs in the observed dSph and the high
presence of them in our cored simulations could be a hint for the
solution of the cusp-core problem, which is a mismatch in the DM halo
profiles between the prediction of cosmological $\Lambda$-CDM
simulations, which say that the DM halos have to follow a cusped
profile, and the observed kinematic data in dSphs which hints to the
notion that the DM profile is possibly cored.  In our models the SCs
were dissolved more quickly in a cusped profile than in a cored DM
halo, this could be a hint of the existence of a mechanism (e.g.\
feedback, modifications of the nature of dark matter) that eliminates
the cusp in DM halos, as some authors claim
\citep[e.g.][]{Governato2010}.  As we are not performing any
sophisticated combined DM and hydrodynamical simulations such
transformations from cusped to cored halos and why they are happening
are beyond the scope of this study.

Finally, we have to keep in mind that our models make a lot of
simplifications, for example, we put the distribution of SCs into
virial equilibrium within the DM halo for simplicity, because we do
not know in which virial state the SCs are formed and also they
follow a Plummer distribution because we expect to form more SCs in
the centre of the DM halo than further out.  Also we assume that all
SCs have a Plummer sphere distribution, which is a good approximation
to model a SC.  Also all SCs have the same mass.  In reality, we
would expect that in dSphs with low a star formation rate (SFR) we
produce more and smaller SCs and with high SFR we would produce less
and more massive SCs.  Furthermore, objects with recent star formation
are also the ones showing low SFRs and therefore the SCs would be easier
to destroy.  So we overestimate the number of surviving SCs in these
simulations. 

We assume the same SFE for each SC, and after the gas-expulsion the
gas is lost to nowhere.  This is justified because the mass of
the gas is negligible with respect to the mass of the DM halo. 

Another simplification of our models is that we use SUPERBOX, a
particle mesh code which do not take in account two body
interactions.  It is useful in our case because a dSph galaxy is a
collision-less system and the SCs are a collisional system only in
their first 4 Myr before gas-expulsion, so we can neglect these
interactions.  

Our previous models \citep{Assmann2013a,Assmann2013b} were subject to
questions, as we do not include any mass-metallicity relation found
in the dwarf galaxy population.  Metallicity information is virtually
impossible to include in a pure stellar dynamical simulation, but with
our new models we show that our previous results are not
altered by including different observed SFHs.  Still our models are
purely stellar dynamical, but now one could include analytically ones
favourite astro-chemical model to derive the yields for the later
generations of SCs and paint the phase-space elements of our
simulations according to their metal content.

In this paper we do not discuss the ellipticities, asymmetries, A4
parameters or what we have dubbed fossil remnants in the velocity
space of our models.  These parameters were discussed in detail in
\citet{Assmann2013b} and the simple addition of a SFH to our models
does not change the findings of our previous manuscript.

\section{Conclusions}
\label{sec:conc}

In this paper we test the possible scenario for the formation of dSph
galaxies proposed by \citet{Assmann2013a,Assmann2013b} by adding
different SFHs to the numerical simulations.  In our simulations we
consider the evolution of 30 star clusters placed at different moments
in time
into our simulations to mimic the SFHs.  The SCs are dissolving within a
cored or a cusped dark matter halo.  Also we study the effect of
different SFEs (20\% and 30\%) for those SCs. 

We observe that after 10~Gyr of evolution we get an object that
resembles the properties of a classical dSph galaxy if we have enough
time to dissolve the SCs. 

In our models with a low SFE=20\% and a DM halo with a NFW profile,
the SCs are dissolved in less than 300~Myr, so we can resemble the
properties of a classical dSph even if we have recent star formation
histories.  Models with a cored DM halo following a Plummer sphere
profile, need more time to dissolve the SCs ($\approx 500$~Myr), and
if the apo-centric distance of these SCs is near to the centre of the DM
halo ($< 300$~pc), they  will need approximately 2~Gyr to get
dissolved.  So we can resemble the properties of a classical dSph only
if we have no recent star formation. 

In our models using SCs with a high SFE=30\% and a DM halo following a
NFW profile, the SCs need approximately between 300~Myr to 1~Gyr to
dissolve, and again as we do not see surviving SCs in classical
dSph.   On the other hand, models with a cored DM profile and SCs with
a SFE=30\%, can not resemble the properties of classical dSph, because
some of the SCs cannot dissolve if their apo-centric distance is
close to the centre of the DM halo, they need more than 10~Gyr
to dissolve completely, but this could be a hint to explain the
formation of massive dSph galaxies like Fornax, with 5 orbiting SCs,
or Sagittarius.  

In all our models even if we have surviving SCs, we obtain surface
brightness profiles as observed in classical dSph.  We use a Sersic
profile to fit the surface brightness profile of our final objects and
get effective radii similar to the observed ones in dSph with a mean
value of $245 \pm 26$~pc for cusped and $523 \pm 195$~pc for cored
profiles, also we get values of the $n$ index close to unity
($n = 0.74 \pm 0.16$ for cusped and $n = 0.84 \pm 0.29$ for cored),
this means that our final objects have exponential surface brightness
profiles, as observed in dSphs. 

Furthermore, all our simulations show a velocity dispersion in the
observed range of classical dSph, between 5-12~km\,s$^{-1}$, and
dispersion profiles which remain flat independent of the radius.
Another important feature is that we
see wiggles and bumps in the data.  In observational papers these
wiggles and bumps are smoothed over but according to our scenario some
of them could be real.  In reality we see similar bumps in Carina,
Leo~I and Sculptor \citep{Walker2009}.  

Our models give a hint to the cusp-core problem, as the SCs are more
likely to get dissolved in a cusped DM halo, and observations hint
that DM halos of dSph could follow a cored profile.  We suspect that
there could be a mechanism that eliminates the cusp after dissolving
all the SCs and form a cored DM profile. 

Our simulations show that our formation scenario works even if using
different SFHs, DM halo profiles and different SFE to dissolve the SCs
and we are not only able to reproduce the observational data that we
have today, but provide observers with predictions for future
observations. \\

{\bf Acknowledgements:}
AA thanks R. Dominguez, B. Reinoso and N. Araneda for their help
with SUPERBOX.  AA thanks his friend N.P. Less for the interesting 
discussion during the realisation of this work.  AA and FU 
acknowledge financial support from Conicyt through the project 
PII20150171.  MF acknowledges financial support through Fondecyt 
grant No.~1130521, Conicyt PII20150171 and BASAL PFB-06/2007. 

\bibliographystyle{mn2e}

\begin{thebibliography}{99}

\bibitem[\protect\citeauthoryear{Aarseth, Henon \& Wielen}{Aarseth et al.}
  {1974}]{Aarseth1974}
Aarseth S.J., Henon M., Wielen R., 1974, A\&A, 37, 183

\bibitem[\protect\citeauthoryear{Amorisco}{2017}]{amorisco2017}
Amorisco N.C., 2017, ApJ, 844, 64 

\bibitem[\protect\citeauthoryear{Assmann et al.}{2013a}]{Assmann2013a}
Assmann P., Fellhauer M., Wilkinson M.I., Smith R., 2013a, MNRAS, 435.2391

\bibitem[\protect\citeauthoryear{Assmann et al.}{2013b}]{Assmann2013b}
Assmann P., Fellhauer M., Wilkinson M.I., Smith R., 2013b, MNRAS, 432,274

\bibitem[\protect\citeauthoryear{Belokurov et al.}{2007}]{Belokurov2007}
Belokurov V. et al., 2007, ApJ, 654, 897

\bibitem[\protect\citeauthoryear{Boily \& Kroupa}{2003a}]{Boily2003a}
Boily C.M., Kroupa P., 2003a, MNRAS, 338, 665

\bibitem[\protect\citeauthoryear{Boily \& Kroupa}{2003b}]{Boily2003b}
Boily C.M., Kroupa P., 2003b, MNRAS, 338, 673

\bibitem[\protect\citeauthoryear{Bonnell et al.}{2011}]{Bonnell2011}
Bonnell I.A., Smith R.J., Clark P.C., Bate M.R., 2011, MNRAS, 410, 2339

\bibitem[\protect\citeauthoryear{Bressert et al.}{2010}]{Bressert2010}
Bressert E. et al., 2010, MNRAS, 409, L54

\bibitem[\protect\citeauthoryear{Caon, Capaccioli \& D'Onofrio}{Caon et al.}
  {1993}]{Caon1993}
Caon N., Capaccioli M., D'Onofrio M., 1993, MNRAS, 265, 10132

\bibitem[\protect\citeauthoryear{Cole et al.}{1994}]{Cole1994}
Cole S., Aragon-Salamanca A., Frenk C.S., Navarro J.F., Zepf
S.E., 1994, MNRAS, 271, 781 

\bibitem[\protect\citeauthoryear{Conselice}{2003}]{Conselice2003}
Conselice C. J., 2003, ApJS, 147, 1

\bibitem[\protect\citeauthoryear{Dehnen \& McLaughlin}{2005}]{Dehnen2005}
Dehnen W., McLaughlin D.E., 2005, MNRAS, 363, 1057

\bibitem[\protect\citeauthoryear{D'Onghia et al.}{2009}]{DOnghia2009}
D Onghia E., Besla G., Cox T., Hernquist L., 2009, Nature, 460, 605

\bibitem[\protect\citeauthoryear{Fellhauer et al.}{2000}]{Fellhauer2000}
Fellhauer M., Kroupa P., Baumgardt H., Bien R., Boily C.M., Spurzem
R., Wassmer N., 2000, New Astron., 5, 305 

\bibitem[\protect\citeauthoryear{Gnedin et al.}{1999}]{Gnedin1999}
Gnedin O.Y., Hernquist L., Ostriker J. P., 1999, ApJ, 514, 109

\bibitem[\protect\citeauthoryear{Goodwin}{1997a}]{Goodwin1997a}
Goodwin S.P., 1997a, MNRAS, 284, 785

\bibitem[\protect\citeauthoryear{Goodwin}{1997b}]{Goodwin1997b}
Goodwin S.P., 1997b, MNRAS, 286, 669

\bibitem[\protect\citeauthoryear{Governato et al.}{2010}]{Governato2010}
Governato F., Brook C., Mayer L., Brooks A., Rhee G., Jonsson P.,
Willman B., Stinson G., Quinn T., Madau P., 2010, Nature, 463,
203

\bibitem[\protect\citeauthoryear{Irwin \& Hatzidimitriou}{1995}]{Irwin1995}
Irwin M., Hatzidimitriou D., 1995, MNRAS, 277, 1354

\bibitem[\protect\citeauthoryear{Jerjen, Binggeli \& Freeman}{Jerjen et al.}
  {2000}]{Jerjen2000}
Jerjen H., Binggeli B., Freeman K.C., 2000, AJ, 119, 593

\bibitem[\protect\citeauthoryear{Kauffmann, White \& Guiderdoni}
  {Kauffmann et al.}{1993}]{Kauffmann1993}
Kauffmann G., White S., Guiderdoni B., 1993, MNRAS, 264,201

\bibitem[\protect\citeauthoryear{Kleyna et al.}{2002}]{Kleyna2002}
Kleyna J.T., Wilkinson M.I., Evans N.W., Gilmore G., Frayn C., 2002,
MNRAS, 330, 792

\bibitem[\protect\citeauthoryear{Kleyna et al.}{2003}]{Kleyna2003}
Kleyna J.T., Wilkinson M.I., Gilmore G., Evans N. W., 2003, ApJ, 588,
L21

\bibitem[\protect\citeauthoryear{Kleyna et al.}{2004}]{Kleyna2004}
Kleyna J.T., Wilkinson M.I., Evans N.W., Gilmore G., 2004, MNRAS,
354, L66

\bibitem[\protect\citeauthoryear{Koch et al.}{2009}]{Koch2009}
Koch A. et al., 2009, ApJ, 690, 453

\bibitem[\protect\citeauthoryear{Lada \& Lada}{2003}]{Lada2003}
Lada C.J., Lada E.A., 2003, ARA\&A, 41, 57

\bibitem[\protect\citeauthoryear{Lada, Lombardi \& Alves}{Lada et al.}
  {2010}]{Lada2010}
Lada C.J., Lombardi M., Alves J.F., 2010, ApJ, 724, 687

\bibitem[\protect\citeauthoryear{Lokas}{2009}]{Lokas2009}
Lokas E.L., 2009, MNRAS, 394, L102

\bibitem[\protect\citeauthoryear{Mayer et al.}{2007}]{Mayer2007}
Mayer L., Kazantzidis S., Mastropiero C., Wadsley J., 2007, Nat, 445, 738

\bibitem[\protect\citeauthoryear{Mateo M.}{1998}]{Mateo1998}
Mateo M.L., 1998, ARA\&A, 36, 435

\bibitem[\protect\citeauthoryear{McConnachie}{2012}]{McConnachie2012}
McConnachie, A.W., 2012, AJ, 1444

\bibitem[\protect\citeauthoryear{Munoz et al.}{2005}]{Munoz2005}
Munoz R.R. et al., 2005, ApJ, 631, L137

\bibitem[\protect\citeauthoryear{Munoz et al.}{2006}]{Munoz2006}
Munoz R.R. et al., 2006, ApJ, 649, 201

\bibitem[\protect\citeauthoryear{Navarro, Frenk \& White}{Navarro et al.}
  {1997}]{Navarro1997}
Navarro J.F., Frank C.S., White S.D.M., 1997, ApJ, 490, 493

\bibitem[\protect\citeauthoryear{Parmentier et al.}{2008}]{Parmentier2008}
  Parmentier G., Goodwin S.P., Kroupa P., Baumgardt H., 2008, ApJ, 678, 347

\bibitem[\protect\citeauthoryear{Pe\~{n}arrubia, Walker \& Gilmore}
  {Pe\~{n}arrubia et al.}{2009}]{pen09}
  Pe\~{n}arrubia J., Walker M.G., Gilmore G., 2009, MNRAS, 399, 1275

\bibitem[\protect\citeauthoryear{Plummer}{1911}]{Plummer1911}
Plummer H.C., 1911, MNRAS, 71, 460

\bibitem[\protect\citeauthoryear{Simon \& Geha}{2007}]{Simon2007}
Simon J.D., Geha M., 2007, ApJ, 670, 313

\bibitem[\protect\citeauthoryear{Grasha et al.}{2017}]{grasha2017}
  Grasha K. et al., 2017, ApJ, 840, 113

\bibitem[\protect\citeauthoryear{Smith et al.}{2011a}]{Smith2011a}
Smith R., Slater R., Fellhauer M., Goodwin S.P., Assmann P., 2011a,
MNRAS, 416, 383 

\bibitem[\protect\citeauthoryear{Smith et al.}{2011b}]{Smith2011b}
Smith R., Fellhauer M., Goodwin S.P., Assmann P., 2011b, MNRAS, 414, 3036

\bibitem[\protect\citeauthoryear{Walcher et al.}{2003}]{Walcher2003}
Walcher C.J., Fried J.W., Burkert A., Klessen R.S., 2003, A\&A, 406, 847

\bibitem[\protect\citeauthoryear{Walker et al.}{2007}]{Walker2007}
Walker M.G., Mateo M., Olszewski E.W., Gnedin O.Y., Wang X.,
Bodhisattva S., Woodroofe M., 2007, ApJ, 667, L53

\bibitem[\protect\citeauthoryear{Walker et al.}{2009}]{Walker2009}
Walker M.G., Mateo M., Olszewski E.W., Penarrubia J., Evans N.W.,
Gilmore G., 2009, ApJ, 704, 1274

\bibitem[\protect\citeauthoryear{Weisz et al.}{2014}]{Weisz2014}
  Weisz D.R., Dolphin A.E., Skillman E.D., Holtzman J., Gilbert K.M.,
  Dalcanton J.J., Williams B.F., 2014, ApJ, 789, 147

\bibitem[\protect\citeauthoryear{Withmore et al.}{1999}]{Withmore1999}
Whitmore B.C., Zhang Q., Leitherer C., Fall S.M., Schweizer F.,
Miller B.W., 1999, AJ, 118, 1551 	
    
\end{thebibliography}

\label{lastpage}

\end{document}